\RequirePackage{ifpdf}
\ifpdf 
\documentclass[pdftex]{sigma}
\else
\documentclass{sigma}
\fi

\numberwithin{equation}{section}

\begin{document}

\allowdisplaybreaks

\renewcommand{\thefootnote}{$\star$}

\renewcommand{\PaperNumber}{008}

\FirstPageHeading

\ShortArticleName{Form Factors of
Belavin's $(\mathbb{Z}/n\mathbb{Z})$-Symmetric Model and
Its Application}

\ArticleName{A Vertex Operator Approach for
Form Factors \\ of
Belavin's $\boldsymbol{(\mathbb{Z}/n\mathbb{Z})}$-Symmetric Model\\ and
Its Application\footnote{This paper is a
contribution to the Proceedings of the International Workshop ``Recent Advances in Quantum Integrable Systems''. The
full collection is available at
\href{http://www.emis.de/journals/SIGMA/RAQIS2010.html}{http://www.emis.de/journals/SIGMA/RAQIS2010.html}}}

\Author{Yas-Hiro QUANO}

\AuthorNameForHeading{Y.-H.~Quano}

\Address{Department of Clinical Engineering,
Suzuka University of Medical Science, \\
Kishioka-cho, Suzuka 510-0293, Japan}
\Email{\href{mailto:quanoy@suzuka-u.ac.jp}{quanoy@suzuka-u.ac.jp}}

\ArticleDates{Received October 22, 2010, in f\/inal form January 07, 2011;  Published online January 15, 2011}

\Abstract{A vertex operator approach for form factors of
Belavin's $(\mathbb{Z}/n\mathbb{Z})$-symmetric model
is constructed
on the basis of bosonization of vertex operators
in the $A^{(1)}_{n-1}$ model and vertex-face transformation.
As simple application for $n=2$, we obtain expressions for
$2m$-point form factors related to
the $\sigma^z$ and $\sigma^x$ operators in
the eight-vertex model.}


\Classification{37K10; 81R12}

\renewcommand{\thefootnote}{\arabic{footnote}}
\setcounter{footnote}{0}

\section{Introduction}

In \cite{Bel-corr} and \cite{Bel-form}
we derived the integral formulae for correlation functions and form
factors, respectively, of
Belavin's $(\mathbb{Z}/n\mathbb{Z})$-symmetric model~\cite{Bela,RT}
on the basis of vertex operator approach~\cite{JMbk}.
Belavin's $(\mathbb{Z}/n\mathbb{Z})$-symmetric model
is an $n$-state generalization of
Baxter's eight-vertex model~\cite{ESM}, which has
$(\mathbb{Z}/2\mathbb{Z})$-symmetries.
As for the eight-vertex model, the integral formulae for correlation functions
and form factors were derived by Lashkevich and Pugai~\cite{LaP} and
by Lashkevich~\cite{La}, respectively.

It was found in \cite{LaP} that
the correlation functions of the eight-vertex model can be obtained by
using the free f\/ield realization of the vertex operators in the
eight-vertex SOS model \cite{LuP}, with insertion of the nonlocal operator
$\Lambda$, called `the tail operator'.
The vertex operator approach for higher spin generalization of the
eight-vertex model was presented in \cite{KKW}.
The vertex operator approach for higher rank generalization was
presented in~\cite{Bel-corr}.
The expression of the spontaneous polarization of
the $(\mathbb{Z}/n\mathbb{Z})$-symmetric model~\cite{SPn} was also reproduced
in \cite{Bel-corr}, on the basis of vertex operator approach.
Concerning form factors, the bosonization scheme for
the eight-vertex model was constructed in~\cite{La}.
The higher rank generalization of~\cite{La} was presented
in~\cite{Bel-form}. It was shown in~\cite{KK1,KK2} that
the elliptic algebra
$U_{q,p}(\widehat{\mathfrak s\mathfrak l}_N)$ relevant to
the $(\mathbb{Z}/n\mathbb{Z})$-symmetric model
provides
the Drinfeld realization of the face type elliptic quantum group
${\cal B}_{q,\lambda}(\widehat{\mathfrak s\mathfrak l}_N)$ tensored by
a Heisenberg algebra.

The present paper is organized as follows. In Section~\ref{section2}
we review the basic def\/initions of
the $(\mathbb{Z}/n\mathbb{Z})$-symmetric model~\cite{Bela},
the corresponding dual face model $A^{(1)}_{n-1}$ model~\cite{JMO},
and the vertex-face correspondence.
In Section~\ref{section3} we summarize the vertex operator algebras
relevant to the $(\mathbb{Z}/n\mathbb{Z})$-symmetric model
and the $A^{(1)}_{n-1}$ model \cite{Bel-corr,Bel-form}.
In Section~\ref{section4} we construct the free f\/ield representations of
the tail operators, in terms of those
of the basic operators for
the type~I~\cite{AJMP} and the type~II~\cite{FKQ}
vertex operators in the $A^{(1)}_{n-1}$ model.
Note that in the present paper we use
a dif\/ferent convention from the one used in
\cite{Bel-corr,Bel-form}. In Section~\ref{section5} we calculate
$2m$-point form factors of the $\sigma^z$-operator
and $\sigma^x$-operator in
the eight-vertex model,
as simple application for $n=2$.
In Section~\ref{section6} we give some concluding remarks.
Useful operator product expansion (OPE) formulae and
commutation relations for basic bosons
are given in Appendix~\ref{AppendixA}.

\section{Basic def\/initions}\label{section2}

The present section aims to formulate
the problem, thereby f\/ixing the notation.

\subsection{Theta functions}

The Jacobi theta function with two pseudo-periods $1$ and
$\tau$ (${\rm Im}\,\tau >0$) are def\/ined as follows:
\begin{gather*}
\vartheta\left[\begin{array}{c} a \\ b \end{array} \right]
(v;\tau ): =\displaystyle\sum_{m\in \mathbb{Z}}
\exp \left\{ \pi \sqrt{-1}(m+a)~\left[ (m+a)\tau
+2(v+b) \right] \right\}, 
\end{gather*}
for $a,b\in\mathbb{R}$.
Let $n\in\mathbb{Z}_{\geqslant 2}$ and
$r\in\mathbb{R}_{>1}$, and also f\/ix
the parameter $x$ such that $0<x<1$.
We will use the abbreviations,
\begin{alignat*}{4}
& [v]:=x^{\frac{v^2}{r}-v}\Theta_{x^{2r}}(x^{2v}),\qquad && [v]':=[v]|_{r\mapsto r-1},\qquad  &&
[v]_1:=[v]|_{r\mapsto 1},  & \\
& \{v\}:=x^{\frac{v^2}{r}-v}\Theta_{x^{2r}}(-x^{2v}), \qquad && \{v\}':=\{v\}|_{r\mapsto r-1}, \qquad &&
\{v\}_1:=\{v\}|_{r\mapsto 1}, &
\end{alignat*}
where
\begin{gather*}
 \Theta_{q}(z)=(z; q)_\infty
\big(qz^{-1}; q\big)_\infty (q; q)_\infty =
\sum_{m\in\mathbb{Z}} q^{m(m-1)/2}(-z)^m, \\
 (z; q_1 , \dots , q_m )_\infty =
\prod_{i_1 , \dots , i_m \geqslant 0}
\big(1-zq_1^{i_1} \cdots q_m^{i_m}\big).
\end{gather*}
Note that
\begin{gather*}
\vartheta\left[\begin{array}{c} 1/2 \\ -1/2 \end{array} \right]
\left( \frac{v}{r}, \frac{\pi\sqrt{-1}}{\epsilon r} \right)
=\sqrt{\frac{\epsilon r}{\pi}}
\exp\,\left(-\frac{\epsilon r}{4}\right)[v],\\
\vartheta\left[\begin{array}{c} 0 \\ 1/2 \end{array} \right]
\left( \frac{v}{r}, \frac{\pi\sqrt{-1}}{\epsilon r} \right)
=\sqrt{\frac{\epsilon r}{\pi}}
\exp\,\left(-\frac{\epsilon r}{4}\right)\{v\},
\end{gather*}
where $x=e^{-\epsilon}$ ($\epsilon >0$).

For later conveniences we also
introduce the following symbols:
\begin{alignat}{3}
& r_{j}(v)=z^{\frac{r-1}{r}\frac{n-j}{n}}
\frac{g_{j}(z^{-1})}{g_{j}(z)}, \qquad &&
g_{j}(z)=
\frac{\{x^{2n+2r-j-1}z\}
\{x^{j+1}z\}}
{\{x^{2n-j+1}z\}\{x^{2r+j-1}z\}}, \label{eq:g-def} & \\
& r^*_{j}(v)=z^{\frac{r}{r-1}\frac{n-j}{n}}
\frac{g^*_{j}(z^{-1})}{g^*_{j}(z)}, \qquad &&
g^*_{j}(z)=
\frac{\{x^{2n+2r-j-1}z\}'
\{x^{j-1}z\}'}
{\{x^{2n-j-1}z\}'\{x^{2r+j-1}z\}'},  \label{eq:g*-def} & \\
& \chi_j (v) = (-z)^{-\frac{j(n-j)}{n}} \dfrac{\rho_j (z^{-1})}{
\rho_j (z)}, \qquad &&
\rho_j (z)=\frac{(x^{2j+1}z;x^2,x^{2n})_\infty
(x^{2n-2j+1}z;x^2,x^{2n})_\infty
}{(xz;x^2,x^{2n})_\infty
(x^{2n+1}z;x^2,x^{2n})_\infty
}, \!\!\!\!\!\!\label{eq:chi-def}&
\end{alignat}
where $z=x^{2v}$, $1\leqslant j\leqslant n$ and
\begin{gather*}
\{z\}=(z;x^{2r},x^{2n})_\infty , \qquad
\{z\}'=(z;x^{2r-2},x^{2n})_\infty .
\end{gather*}
In particular we denote $\chi (v)=\chi_1 (v)$.
These factors will appear in the commutation relations
among the type I and type II vertex operators.

The integral kernel for the type I and the type II
vertex operators will be given as the products of
the following elliptic functions:
\begin{alignat*}{3}
& f(v,w)=\frac{[v+\frac{1}{2}-w]}{[v-\frac{1}{2}]},\qquad && h(v)=\frac{[v-1]}{[v+1]}, & \\ 
& f^*(v,w)=\frac{[v-\frac{1}{2}+w]'}{[v+\frac{1}{2}]'},\qquad &&
h^*(v)=\frac{[v+1]'}{[v-1]'}.& 
\end{alignat*}

\subsection[Belavin's $(\mathbb{Z}/n\mathbb{Z})$-symmetric model]{Belavin's $\boldsymbol{(\mathbb{Z}/n\mathbb{Z})}$-symmetric model}

Let $V=\mathbb{C}^n$ and
$\{ \varepsilon _\mu \}_{0 \leqslant \mu \leqslant n-1}$ be
the standard orthonormal basis with the inner
product $\langle \varepsilon _\mu ,
\varepsilon _\nu \rangle =\delta_{\mu \nu}$.
Belavin's $(\mathbb{Z}/n\mathbb{Z})$-symmetric model~\cite{Bela} is
a vertex model on a two-dimensional square lattice ${\cal L}$
such that the state variables take the values of
$(\mathbb{Z}/n\mathbb{Z})$-spin. The model is
$(\mathbb{Z}/n\mathbb{Z})$-symmetric in a sense that the $R$-matrix satisf\/ies
the following conditions:
\begin{alignat*}{3}
& \mbox{({\romannumeral 1})} \quad & & R(v)^{ik}_{jl}=0, \mbox{~~unless $i+k=j+l$,~~mod $n$},&  \\
& \mbox{({\romannumeral 2})} \quad & & R(v)^{i+p k+p}_{j+p l+p}= R(v)^{ik}_{jl}, \mbox{~$\forall\, i,j,k,l,p\in \mathbb{Z}/n\mathbb{Z}$}.&
\end{alignat*}
The def\/inition of the
$R$-matrix in the principal regime can be found in \cite{Bel-form}.
The present $R$-matrix has three parameters
$v$, $\epsilon$ and $r$, which lie
in the following region:
\begin{gather*}
\epsilon >0, \qquad r>1, \qquad 0<v<1.
\end{gather*}

\subsection[The $A^{(1)}_{n-1}$ model]{The $\boldsymbol{A^{(1)}_{n-1}}$ model}

The dual face model of the $(\mathbb{Z}/n\mathbb{Z})$-symmetric model
is called the $A^{(1)}_{n-1}$ model. This is a face model
on a two-dimensional square lattice ${\cal L^*}$, the dual lattice
of ${\cal L}$,
such that the state variables take the values of
the dual space of Cartan subalgebra ${\mathfrak h}^*$ of $A^{(1)}_{n-1}$:
\begin{gather*}
{\mathfrak h}^*=\bigoplus_{\mu =0}^{n-1}
\mathbb{C} \omega_\mu ,
\end{gather*}
where
\begin{gather*}
\omega_\mu :=\sum_{\nu =0}^{\mu -1} \bar{\varepsilon}_\nu ,
\qquad
\bar{\varepsilon}_\mu =
\varepsilon _\mu -\frac{1}{n}\sum_{\mu =0}^{n-1}
\varepsilon _\mu .
\end{gather*}
The weight lattice $P$ and the root lattice $Q$ of
$A^{(1)}_{n-1}$ are usually def\/ined.
For $a\in {\mathfrak h}^*$, we set
\begin{gather*}
a_{\mu\nu}=\bar{a}_\mu-\bar{a}_\nu , \qquad
\bar{a}_\mu =\langle a+\rho ,
\varepsilon_\mu \rangle =\langle a+\rho ,
\bar{\varepsilon}_\mu \rangle , \qquad
\rho =\sum_{\mu =1}^{n-1} \omega_\mu .
\end{gather*}

An ordered pair $(a,b) \in {\mathfrak h}^{*2}$
is called admissible if $b=a+\bar{\varepsilon}_\mu$,
for a certain $\mu\,(0\leqslant \mu \leqslant n-1)$.
For $(a, b, c, d)\in {\mathfrak h}^{*4}$, let
$\displaystyle W
\left[ \left. \begin{array}{cc}
c & d \\ b & a \end{array}
\right| v \right] $
be the Boltzmann weight of the
$A^{(1)}_{n-1}$ model for the state conf\/iguration
$\displaystyle
\left[ \begin{array}{cc}
c & d \\ b & a \end{array} \right] $
round a face.
Here the four states $a$, $b$, $c$ and $d$ are
ordered clockwise from the SE corner.
In this model $W
\left[ \left. \begin{array}{cc}
c & d \\ b & a \end{array} \right|
v \right] =0~~$
unless the four pairs $(a,b), (a,d), (b,c)$
and $(d,c)$ are admissible. Non-zero Boltzmann weights
are parametrized in terms of
the elliptic theta function of the spectral parameter~$v$.
The explicit expressions of~$W$ can be found in~\cite{Bel-form}.
We consider the so-called Regime III in the model, i.e.,
$0<v<1$.

\subsection{Vertex-face correspondence}

Let $t(v)^a_{a-\bar{\varepsilon}_\mu}$
be the intertwining vectors in $\mathbb{C}^n$,
whose elements are expressed in terms of theta functions.
As for the def\/initions see \cite{Bel-form}.
Then $t(v)^a_{a-\bar{\varepsilon}_\mu}$'s relate
the $R$-matrix of
the $(\mathbb{Z}/n\mathbb{Z})$-symmetric model in
the principal regime and Boltzmann weights $W$ of the
$A^{(1)}_{n-1}$ model in the regime~III
\begin{gather}
R(v_1-v_2)t (v_1)_a^d\otimes t (v_2)_d^c=
\sum_{b} t(v_1)_b^c \otimes t (v_2)_a^b
W\left[ \left.
\begin{array}{cc} c & d \\ b & a \end{array} \right|
v_1 -v_2 \right].
\label{eq:Rtt=Wtt}
\end{gather}

Let us introduce the dual intertwining vectors
satisfying
\begin{gather}
\sum_{\mu =0}^{n-1} t_\mu^*  (v)^{a'}_{a}
t^\mu (v)^{a}_{a''} =\delta_{a''}^{a'}, \qquad
\sum_{\nu =0}^{n-1} t^\mu (v)^{a}_{a-\bar{\varepsilon}_\nu}
t_{\mu'}^* (v)^{a-\bar{\varepsilon}_\nu}_{a} =
\delta^\mu_{\mu'}. \label{eq:dual-t}
\end{gather}

From (\ref{eq:Rtt=Wtt}) and (\ref{eq:dual-t}), we have
\begin{gather*}
t^*(v_{1})^{b}_{c}\otimes t^*(v_{2})^{a}_{b}
R(v_{1}-v_2 )=
 \sum_{d}
W\left[ \left. \begin{array}{cc}
c & d \\ b & a \end{array} \right| v_{1}-v_2 \right]
t^*(v_{1} )^{a}_{d}\otimes t^*(v_{2} )^{d}_{c}.
\end{gather*}

For f\/ixed $r>1$, let
\begin{gather*}
S(v )=-R(v)|_{r\mapsto r-1}, \qquad
W'\left[ \left.
\begin{array}{cc} c & d \\ b & a \end{array} \right|
v \right]=-W\left[ \left.
\begin{array}{cc} c & d \\ b & a \end{array} \right|
v \right] \left. \makebox{\rule[-4mm]{0pt}{11mm}}
\right|_{r\mapsto r-1},
\end{gather*}
and $t'{}^* (v)^{b}_{a}$ is the dual intertwining vector
of $t' (v)^{a}_{b}$. Here,
\begin{gather*}
t' (v)^{a}_{b}:=f'(v) t (v; \epsilon , r-1)_{b}^{a},
\end{gather*}
with
\begin{gather}
f'(v)=\dfrac{x^{-\tfrac{v^2}{n(r-1)}-\tfrac{(r+n-2)v}{n(r-1)}-
\tfrac{(n-1)(3r+n-5)}{6n(r-1)}}}{
\sqrt[n]{-(x^{2r-2}; x^{2r-2})_\infty}} \nonumber\\
\hphantom{f'(v)=}{}\times \dfrac{
(x^2z^{-1}; x^{2n}, x^{2r-2})_\infty (x^{2r+2n-2}z; x^{2n}, x^{2r-2})_\infty}{
(z^{-1}; x^{2n}, x^{2r-2})_\infty (x^{2r+2n-4}z; x^{2n}, x^{2r-2})_\infty},
\label{eq:f'-def}
\end{gather}
and $z=x^{2v}$. Then we have
\begin{gather*}
 t'{}^*(v_1 )^{b}_{c}\otimes
t'{}^*(v_2 )^{a}_{b}S(v_1 -v_2 )
=\displaystyle\sum_{d}
W'\left[ \left. \begin{array}{cc}
c & d \\ b & a \end{array} \right| v_1 -v_2 \right]
t'{}^*(v_1 )_{d}^{a}\otimes t'{}^*(v_2 )_{c}^{d}.
\end{gather*}

\section{Vertex operator algebra}\label{section3}

\subsection{Vertex operators for the $(\mathbb{Z}/n\mathbb{Z})$-symmetric model}

Let ${\cal H}^{(i)}$ be the $\mathbb{C}$-vector space
spanned by the half-inf\/inite pure tensor vectors of the forms
\begin{gather*}
\varepsilon_{\mu_1}\otimes \varepsilon_{\mu_2}\otimes
\varepsilon_{\mu_3}\otimes \cdots
\qquad \mbox{with $\mu_j\in \mathbb{Z}/n\mathbb{Z}$,
$\mu_j=i+1-j$ (mod $n$) for $j\gg 0$}.
\end{gather*}
The type I vertex operator $\Phi^\mu (v)$ can be def\/ined as a
half-inf\/inite transfer matrix.
The opera\-tor~$\Phi^\mu (v)$ is an intertwiner
from ${\cal H}^{(i)}$ to ${\cal H}^{(i+1)}$,
satisfying the following
commutation relation:
\begin{gather*}
\Phi^\mu (v_1)\Phi^\nu (v_2)=
\sum_{\mu',\nu'} R(v_1-v_2)^{\mu\nu}_{\mu'\nu'}
\Phi^{\nu'} (v_2)\Phi^{\mu'} (v_1).
\end{gather*}

When we consider an operator related to `creation-annihilation' process,
we need another type of vertex operators, the type II vertex operators
that satisfy the following commutation relations:
\begin{gather*}
\Psi^*_\nu (v_2)\Psi^*_\mu (v_1)=
\sum_{\mu',\nu'} \Psi^*_{\mu'} (v_1)\Psi^*_{\nu'} (v_2)
S(v_1-v_2)_{\mu\nu}^{\mu'\nu'},
\\
\Phi^\mu (v_1)\Psi^*_\nu (v_2)=\chi (v_1 -v_2)
\Psi^*_{\nu} (v_2)\Phi^{\mu} (v_1).
\end{gather*}

Let
\begin{gather*}
\rho^{(i)}=x^{2nH_{\rm CTM}}: {\cal H}^{(i)}\rightarrow
{\cal H}^{(i)},
\end{gather*}
where $H_{\rm CTM}$ is the CTM Hamiltonian def\/ined as follows:
\begin{gather}
H_{\rm CTM}(\mu_1 , \mu_2 , \mu_3, \dots )=\dfrac{1}{n}
\sum_{j=1}^\infty jH_v (\mu_j , \mu_{j+1}), \nonumber\\
H_v (\mu ,\nu ) = \left\{ \begin{array}{ll}
\mu -\nu -1 & \mbox{if $0\leqslant \nu <\mu\leqslant n-1$}, \\
n-1+\mu -\nu & \mbox{if $0\leqslant \mu\leqslant \nu\leqslant n-1$}.
\end{array} \right.
\label{eq:H_v}
\end{gather}
Then we have the homogeneity relations
\begin{gather*}
\Phi^\mu (v) \rho^{(i)} =\rho^{(i+1)}\Phi^\mu (v-n), \qquad
\Psi^*_\mu (v) \rho^{(i)} =\rho^{(i+1)}\Psi^*_\mu (v-n).
\end{gather*}

\subsection[Vertex operators for the $A^{(1)}_{n-1}$ model]{Vertex operators for the $\boldsymbol{A^{(1)}_{n-1}}$ model}

For $k=a+\rho , l=\xi +\rho$ and $0\leqslant i\leqslant n-1$,
let ${\cal H}^{(i)}_{l,k}$ be the space of admissible paths
$(a_0 , a_1, a_2, \dots )$ such that
\begin{gather*}
a_0 =a, \qquad  a_{j} -a_{j+1}\in \left\{
\bar{\varepsilon}_0 , \bar{\varepsilon}_1 ,
\dots , \bar{\varepsilon}_{n-1}
\right\} \quad \mbox{for $j=0, 1, 2, 3, \dots$,}\\
a_j=\xi +\omega_{i+1-j} \quad \mbox{for $j\gg 0$}.
\end{gather*}
The type I vertex operator $\Phi(v)_a^{a+\bar{\varepsilon}_\mu}$
can be def\/ined as a
half-inf\/inite transfer matrix.
The operator $\Phi(v)_a^{a+\bar{\varepsilon}_\mu}$
is an intertwiner from ${\cal H}^{(i)}_{l,k}$ to
${\cal H}^{(i+1)}_{l,k+\bar{\varepsilon}_\mu}$,
satisfying the following
commutation relation:
\begin{gather*}
\Phi (v_1)^c_b\Phi (v_2)^b_a=
\sum_{d} W\left[ \left. \begin{array}{cc}
c & d \\
b & a \end{array} \right| v_1-v_2 \right]
\Phi (v_2)^{c}_d\Phi (v_1)^{d}_a .
\end{gather*}
The free f\/ield realization of $\Phi (v_2)^b_a$ was constructed
in \cite{AJMP}. See Section~\ref{section4.2}.

The type II vertex operators should satisfy
the following commutation relations:
\begin{gather*}
\Psi^* (v_2)^{\xi_c}_{\xi_d}\Psi^* (v_1)^{\xi_d}_{\xi_a}=
\sum_{\xi_b}
\Psi^* (v_1)^{\xi_c}_{\xi_b}\Psi^* (v_2)^{\xi_b}_{\xi_a}
W'\left[ \left. \begin{array}{cc}
\xi_c & \xi_d \\
\xi_b & \xi_a \end{array} \right| v_1-v_2 \right],
\\
\Phi (v_1)^{a'}_a\Psi^* (v_2)^{\xi'}_{\xi}=
\chi (v_1-v_2)\Psi^* (v_2)^{\xi'}_{\xi}\Phi (v_1)^{a'}_a .
\end{gather*}

Let
\begin{gather*}
\rho^{(i)}_{l,k}=G_a x^{2nH_{l,k}^{(i)}}, \qquad
G_a =\prod_{0\leqslant\mu <\nu\leqslant n-1} [a_{\mu\nu}],
\end{gather*}
where $H_{l,k}^{(i)}$ is the
CTM Hamiltonian of $A^{(1)}_{n-1}$ model
in regime III is given as follows:
\begin{gather*}
H_{l,k}^{(i)}(a_0 , a_1 , a_2, \dots )=\dfrac{1}{n}
\displaystyle\sum_{j=1}^\infty jH_f (a_{j-1} , a_{j}, a_{j+1}), \\
H_f(a+\bar{\varepsilon }_\mu +\bar{\varepsilon }_\nu ,
a+\bar{\varepsilon }_\mu , a )=
H_v (\nu , \mu ),
\end{gather*}
and $H_v (\nu , \mu )$ is the same one as (\ref{eq:H_v}).
Then we have the homogeneity relations
\begin{gather*}
\Phi (v)^{a'}_a \dfrac{\rho^{(i)}_{a+\rho , l}}{G_a}
=\dfrac{\rho^{(i+1)}_{a'+\rho , l}}{G_{a'}}\Phi (v-n)^{a'}_a,
\qquad
\Psi^* (v)^{\xi'}_{\xi} \rho^{(i)}_{k,\xi +\rho} =
\rho^{(i+1)}_{k,\xi' +\rho}\Psi^* (v-n)^{\xi'}_{\xi}.
\end{gather*}
The free f\/ield realization of $\Psi^* (v)^{\xi'}_{\xi}$ was constructed
in \cite{FKQ}. See Section~\ref{section4.3}.

\subsection{Tail operators and commutation relations}

In \cite{Bel-corr} we introduced the intertwining operators between
${\cal H}^{(i)}$ and
${\cal H}^{(i)}_{l,k}$ ($k=l+\omega_{i}$ (mod $Q$)):
\begin{gather*}
T(u ){}^{\xi a_0}=
\prod_{j=0}^\infty
t^{\mu_j}(-u ){}^{a_j}_{a_{j+1}}:
{\cal H}^{(i)}\rightarrow {\cal H}^{(i)}_{l,k}, \\
T(u ){}_{\xi a_0}=
\prod_{j=0}^\infty
t^*_{\mu_j}(-u ){}_{a_j}^{a_{j+1}}:
{\cal H}^{(i)}_{l,k}\rightarrow {\cal H}^{(i)},
\end{gather*}
which satisfy
\begin{gather}
\rho^{(i)}=\left( \dfrac{(x^{2r-2};x^{2r-2})_\infty}{
(x^{2r};x^{2r})_\infty} \right)^{(n-1)(n-2)/2}\dfrac{1}{
G'_\xi} \sum_{k\equiv l+\omega_i\atop\mbox{\scriptsize (mod $Q$)}}
T (u)_{a\xi} \rho^{(i)}_{l,k} T(u)^{a\xi}.
\label{eq:rho-rel}
\end{gather}

In order to obtain the form factors of
the $(\mathbb{Z}/n\mathbb{Z})$-symmetric model, we need
the free f\/ield representations of the tail operator
which is of\/fdiagonal with respect to the boundary conditions:
\begin{gather}
\Lambda (u )_{\xi\,a}^{\xi'a'}=T(u )^{\xi' a'}T(u)_{\xi\, a}:
{\cal H}^{(i)}_{l,k}\rightarrow {\cal H}^{(i)}_{l'k'},
\label{eq:L=TT'}
\end{gather}
where $k=a+\rho$, $l=\xi +\rho$, $k'=a'+\rho$, and $l'=\xi' +\rho$.
Let
\begin{gather*}
L\left[  \left. \begin{array}{cc} a'_0 & a'_1 \\
a_0 & a_1 \end{array} \right| u \right] :=
\sum_{\mu =0}^{n-1} t^*_\mu (-u)_{a_0}^{a_1}
t^\mu (-u)^{a'_0}_{a'_1}.
\end{gather*}
Then we have
\begin{gather*}
\Lambda(u ){}_{\xi\,a_0}^{\xi'a'_0}=
\prod_{j=0}^\infty L\left[  \left. \begin{array}{cc}
a'_j & a'_{j+1} \\
a_j & a_{j+1} \end{array} \right| u \right].
\end{gather*}

From the invertibility of the intertwining vector and its
dual vector, we have
\begin{gather}
\Lambda (u_0)_{\xi\,a}^{\xi'\,a} =\delta_\xi^{\xi'}.
\label{eq:Lamda-delta}
\end{gather}
Note that
the tail operator (\ref{eq:L=TT'}) satisf\/ies
the following intertwining relations \cite{Bel-corr,Bel-form}:
\begin{gather}
\Lambda (u )^{\xi'c}_{\xi\,b}\Phi (v)^b_a=
\sum_{d}L\left[ \left. \begin{array}{cc} c & d \\
b & a \end{array} \right| u-v \right]
\Phi (v)^c_d\Lambda (u)^{\xi'd}_{\xi\,a},
\label{eq:Lambda-phi}
\\
\Psi^* (v)^{\xi_c}_{\xi_d}\Lambda (u )^{\xi_d\,a'}_{\xi_a\,a}=
\sum_{\xi_b}L'\left[ \left. \begin{array}{cc} \xi_c & \xi_d \\
\xi_b & \xi_a \end{array} \right| u+\Delta u-v \right]
\Lambda (u)^{\xi_c\,a'}_{\xi_b\,a}\Psi^* (v)^{\xi_b}_{\xi_a},
\label{eq:Lambda-psi}
\end{gather}
where
\begin{gather*}
L'\left[ \left. \begin{array}{cc} \xi_c & \xi_d \\
\xi_b & \xi_a \end{array} \right| u\right]
=L\left. \left[ \left. \begin{array}{cc} \xi_c & \xi_d \\
\xi_b & \xi_a \end{array} \right| u\right]\right|_{r\mapsto r-1}.
\end{gather*}
We should f\/ind a representation of $\Lambda (u )^{\xi'a'}_{\xi\,a}$
and f\/ix the constant $\Delta u$
that solves (\ref{eq:Lambda-phi}) and (\ref{eq:Lambda-psi}).

\section{Free f\/iled realization}\label{section4}

\subsection{Bosons}

In \cite{FL,AKOS}
the bosons
$B_m^j\,(1\leqslant j \leqslant n-1, m \in \mathbb{Z}
\backslash \{0\})$ relevant to elliptic algebra
were introduced.
For $\alpha, \beta \in {\mathfrak h}^*$ we denote
the zero mode operators by $P_\alpha$, $Q_\beta$.
Concerning commutation relations among these operators
see \cite{FL,AKOS,Bel-form}.

We will deal with the bosonic Fock spaces
${\cal{F}}_{l,k}$, $(l,k \in {\mathfrak h}^*)$
generated by $B_{-m}^j (m>0)$
over the vacuum vectors $|l,k\rangle$ :
\begin{gather*}
{\cal{F}}_{l,k}=
\mathbb{C}[\{ B_{-1}^j, B_{-2}^j,\dots \}_{
1\leqslant j \leqslant n}]|l,k\rangle,
\end{gather*}
where
\begin{gather*}
|l,k\rangle = \exp \left(\sqrt{-1}(\beta_1Q_k+
\beta_2Q_l)\right)|0,0\rangle,
\end{gather*}
and
\begin{gather*}
t^2 -\beta_0 t-1=(t-\beta_1)(t-\beta_2), \qquad
\beta_0 =\dfrac{1}{\sqrt{r(r-1)}}, \qquad \beta_1<\beta_2.
\end{gather*}

\subsection{Type I vertex operators}\label{section4.2}

Let us def\/ine the basic operators for $j=1,\dots,n-1$
\begin{gather*}
U_{-\alpha_j}(v) = z^{\frac{r-1}{r}}
:\exp\left(-\beta_1 \left(\sqrt{-1}Q_{\alpha_j}
+P_{\alpha_j}\log  z)\right)
+\sum_{m \neq 0}\frac{B_m^j-B_m^{j+1}}{m}
(x^jz)^{-m}\right):, \\
U_{\omega_j}(v) = z^{\frac{r-1}{2r}\frac{j(n-j)}{n}}
:\exp\left(\beta_1 \left(\sqrt{-1}Q_{\omega_j}
+P_{\omega_j}\log z)\right)
-\sum_{m\neq 0}\frac{1}{m}\sum_{k=1}^j
x^{(j-2k+1)m} B_m^kz^{-m}\right):,
\end{gather*}
where $\beta_1 =-\sqrt{\frac{r-1}{r}}$ and $z=x^{2v}$ as usual.
The normal product operation places $P_\alpha$'s to
the right of $Q_\beta$'s, as well as $B_m$'s ($m>0$) to
the right of $B_{-m}$'s.
For some useful OPE formulae and commutation relations,
see Appendix~\ref{AppendixA}.

In what follows we set
\begin{gather*}
\pi_\mu=\sqrt{r(r-1)}P_{\bar{\varepsilon}_\mu}, \qquad
\pi_{\mu \nu}=\pi_\mu-\pi_\nu =rL_{\mu\nu}-(r-1)K_{\mu\nu}.
\end{gather*}
The operators $K_{\mu\nu}$, $L_{\mu\nu}$ and $\pi_{\mu \nu}$
act on ${\cal{F}}_{l,k}$ as scalars
$\langle \varepsilon_\mu -\varepsilon_\nu, k\rangle$,
$\langle \varepsilon_\mu -\varepsilon_\nu, l\rangle$ and
$\langle \varepsilon_\mu -\varepsilon_\nu, rl-(r-1)k\rangle$,
respectively. In what follows we often use the symbols
\begin{gather*}
G_K =\prod_{0\leqslant\mu <\nu\leqslant n-1} [K_{\mu\nu}], \qquad
G'_L =\prod_{0\leqslant\mu <\nu\leqslant n-1} [L_{\mu\nu}]'.
\end{gather*}

For $0 \leqslant \mu \leqslant n-1$ the type I vertex operator
$\Phi (v )^{a+\bar{\varepsilon}_\mu}_a$ can be expressed in terms
of $U_{\omega_j}(v)$ and $U_{-\alpha_j}(v)$
on the bosonic Fock space ${\cal{F}}_{l,a+\rho}$. The explicit expression
of $\Phi (v )^{a+\bar{\varepsilon}_\mu}_a$ can found in~\cite{AJMP}.

\subsection{Type II vertex operators}\label{section4.3}

Let us def\/ine the basic operators for $j=1,\dots,n-1$
\begin{gather*}
V_{-\alpha_j}(v) = (-z)^{\frac{r}{r-1}}
:\exp\left(-\beta_2 \left(\sqrt{-1}Q_{\alpha_j}
+P_{\alpha_j}\log  (-z)\right)
-\sum_{m \neq 0}\frac{A_m^j-A_m^{j+1}}{m}
(x^jz)^{-m}\right):, \\
V_{\omega_j}(v) = (-z)^{\frac{r}{2(r-1)}\frac{j(n-j)}{n}}
\nonumber \\
\phantom{V_{\omega_j}(v) =}{}  \times :\exp\left(\beta_2 \left(\sqrt{-1}Q_{\omega_j}
+P_{\omega_j}\log (-z)\right)
+\sum_{m\neq 0}\frac{1}{m} \sum_{k=1}^j
x^{(j-2k+1)m}A_m^kz^{-m}\right):,
\end{gather*}
where $\beta_2 =\sqrt{\frac{r}{r-1}}$ and $z=x^{2v}$, and
$A_m^j=\frac{[rm]_x}{[(r-1)m]_x}B_m^j$.
For some useful OPE formulae and commutation relations, see
Appendix~\ref{AppendixA}.

For $0 \leqslant \mu \leqslant n-1$ the type II vertex operator
$\Psi^* (v )^{\xi+\bar{\varepsilon}_\mu}_\xi$ can be expressed in terms
of $V_{\omega_j}(v)$ and $V_{-\alpha_j}(v)$
on the bosonic Fock space ${\cal{F}}_{\xi +\rho,k}$.
The explicit expression
of $\Psi^* (v )^{\xi+\bar{\varepsilon}_\mu}_\xi$ can found in~\cite{FKQ}.

\subsection{Free f\/ield realization of tail operators}

In order to construct free f\/ield realization of the
tail operators, we also need another type of
basic operators:
\begin{gather*}
W_{-\alpha_j} (v) = ((-1)^rz)^{\frac{1}{r(r-1)}}\\
\phantom{W_{-\alpha_j} (v) =}{}  \times :\exp\left(-\beta_0 \left(\sqrt{-1}Q_{\alpha_j}
+P_{\alpha_j}\log  (-1)^rz)\right)
-\sum_{m \neq 0}\frac{O_m^j-O_m^{j+1}}{m}
(x^jz)^{-m}\right):,
\end{gather*}
where $\beta_0 =\beta_1 +\beta_2 =\tfrac{1}{\sqrt{r(r-1)}}$,
$(-1)^r:=\exp (\pi\sqrt{-1}r)$ and
$O_m^j =\tfrac{[m]_x}{[(r-1)m]_x} B_m^j$. Concerning
useful OPE formulae and commutation relations,
see Appendix~\ref{AppendixA}.

We cite the results on the free f\/ield realization of tail operators.
In \cite{Bel-corr} we obtained the free f\/ield representation of
$\Lambda (u)_{\xi\,a}^{\xi\,a'}$ satisfying
(\ref{eq:Lambda-phi}) for $\xi'=\xi$:
\begin{gather}
\Lambda (u)^{\xi a-\bar{\varepsilon}_{\mu}}_{\xi a-\bar{\varepsilon}_{\nu}}
=  G_K
\oint \prod_{j=\mu +1}^{\nu}
\dfrac{dz_j}{2\pi\sqrt{-1}z_j}
U_{-\alpha_{\mu +1}}(v_{\mu +1}) \cdots
U_{-\alpha_{\nu}}(v_{\nu})\nonumber \\
\phantom{\Lambda (u)^{\xi a-\bar{\varepsilon}_{\mu}}_{\xi a-\bar{\varepsilon}_{\nu}}=}{}
 \times \prod_{j=\mu}^{\nu -1}
(-1)^{L_{j\nu}-K_{j\nu}}f(v_{j+1}-v_j, \pi_{j\nu})
G_K^{-1},
\label{eq:Bose-Lambda}
\end{gather}
where $v_\mu =u$ and $\mu <\nu$.
In \cite{Bel-form} we obtained the free f\/ield representation of
$\Lambda (v)^{\xi +\bar{\varepsilon}_{n-1} a+\bar{\varepsilon}_{n-1}}_{
\xi +\bar{\varepsilon}_{\mu} a+\bar{\varepsilon}_{n-2}}$ satisfying
(\ref{eq:Lambda-psi})
as follows:
\begin{gather}
 \Lambda (u)^{\xi +\bar{\varepsilon}_{n-1} a+\bar{\varepsilon}_{n-1}}_{
\xi +\bar{\varepsilon}_{\mu} a+\bar{\varepsilon}_{n-2}}=
\dfrac{(-1)^{n-\mu}[a_{n-2\,n-1}]}{(x^{-1}-x)(x^{2r}; x^{2r})_\infty^3}
\dfrac{[\xi_{\mu \, n-1}-1]'}{[1]'} G_K G'_L{}^{-1} \nonumber\\
\phantom{\Lambda (u)^{\xi +\bar{\varepsilon}_{n-1} a+\bar{\varepsilon}_{n-1}}_{
\xi +\bar{\varepsilon}_{\mu} a+\bar{\varepsilon}_{n-2}}=}{}
\times
\oint_{C'} \! \prod_{j=\mu +1}^{n-2}\!\frac{dz_j}{2\pi \sqrt{-1} z_j}
W_{-\alpha_{n-1}} \! \left( u-\tfrac{r-1}{2} \right)
V_{-\alpha_{n-2}}(v_{n-2}) \cdots V_{-\alpha_{\mu +1}}(v_{\mu +1})\nonumber \\
\phantom{\Lambda (u)^{\xi +\bar{\varepsilon}_{n-1} a+\bar{\varepsilon}_{n-1}}_{
\xi +\bar{\varepsilon}_{\mu} a+\bar{\varepsilon}_{n-2}}=}{}
\times
\prod_{j=\mu +1}^{n-2} (-1)^{L_{\mu j}-K_{\mu j}}
f^*(v_{j}-v_{j+1},\pi_{\mu j})G_K^{-1}G'_L ,
\label{eq:Lambda-repII}
\end{gather}
for $0\leqslant \mu\leqslant n-2$
with $\Delta u=-\frac{n-1}{2}$ and
$v_{n-1}=u$.
Concerning other types of tail operators $\Lambda (u)_{\xi a}^{\xi a'}$,
the expressions of the free f\/ield representation can be found in~\cite{Bel-corr,Bel-form}.

\subsection{Free f\/ield realization of CTM Hamiltonian}

Let
\begin{gather}
H_F = \sum_{m=1}^\infty
\dfrac{[rm]_x}{[(r-1)m]_x}
\sum_{j=1}^{n-1}\sum_{k=1}^j x^{(2k-2j-1)m}
B_{-m}^k (B_m^j -B_m^{j+1}) +\dfrac{1}{2}
\sum_{j=1}^{n-1} P_{\omega_j}P_{\alpha_j} \nonumber\\
\phantom{H_F}{}
 = \sum_{m=1}^\infty
\dfrac{[rm]_x}{[(r-1)m]_x}
\sum_{j=1}^{n-1}\sum_{k=1}^j x^{(2j-2k-1)m}
(B_{-m}^j -B_{-m}^{j+1}) B_{m}^k +\dfrac{1}{2}
\sum_{j=1}^{n-1} P_{\omega_j}P_{\alpha_j}
\label{eq:CTM-Fock}
\end{gather}
be the CTM Hamiltonian on the Fock space
${\cal F}_{l,k}$ \cite{FHSW}.
Then we have the homogeneity relation
\begin{gather*}
\phi_\mu (z) q^{H_F} =q^{H_F}\phi_\mu \big(q^{-1}z\big),
\end{gather*}
and the trace formula
\begin{gather*}
\mbox{tr}_{{\cal F}_{l,k}}  \left( x^{2n H_F} G_a  \right)
=\dfrac{x^{n |\beta_1 k+\beta_2l|^2}}
{(x^{2n};x^{2n})^{n-1}_\infty}G_a.
\end{gather*}
Let $\rho^{(i)}_{l,k}=G_a x^{2nH_F}$. Then
the relation (\ref{eq:rho-rel}) holds. We thus indentify
$H_F$ with free f\/ield representations of $H_{l,k}^{(i)}$, the
CTM Hamiltonian of $A^{(1)}_{n-1}$ model
in regime~III.

\section[Form factors for $n=2$]{Form factors for $\boldsymbol{n=2}$}\label{section5}

In this section we would like to f\/ind explicit expressions
of form factors for $n=2$ case, i.e., the eight-vertex model
form factors. Here, we adopt the convention that
the components $0$ and $1$ for $n=2$ are denoted by~$+$ and~$-$.
Form factors of the eight-vertex model are
def\/ined as matrix elements of some local operators.
For simplicity, we choose $\sigma^{z}$
as a local operator:
\begin{gather*}
\sigma^{z}=E^{(1)}_{++} -E^{(1)}_{--},
\end{gather*}
where $E^{(j)}_{\mu \mu'}$
is the matrix unit on the $j$-th site. The free f\/ield representation
of $\sigma^{z}$ is given by
\begin{gather*}
\widehat{\sigma^{z}}=\sum_{\varepsilon =\pm} \varepsilon
\Phi^{*}_{\varepsilon} (u)\Phi^{\varepsilon} (u).
\end{gather*}
Here, $\Phi^{*}_{\varepsilon} (u)$ is the dual type~I vertex operator,
whose free f\/iled representation can be found in~\cite{Bel-corr,Bel-form}.

The corresponding form factors
with $2m$ `charged' particles are given by
\begin{gather}
F^{(i)}_{m}(\sigma^{z}; u_1 , \dots , u_{2m})_{
\nu_{1} \cdots \nu_{2m}}=\dfrac{1}{\chi^{(i)}} \mbox{Tr}_{
{\cal H}^{(i)}}  \left( \Psi^{*}_{\nu_{1}} (u_{1})
\cdots \Psi^{*}_{\nu_{2m}} (u_{2m} )
\widehat{\sigma^{z}} \rho^{(i)} \right),
\label{eq:d-ff}
\end{gather}
where
\begin{gather*}
\chi^{(i)}=\mbox{Tr}_{{\cal H}^{(i)}}  \rho^{(i)}
=\dfrac{(x^{4}; x^{4})_\infty}{
(x^{2}; x^{2})_\infty}.
\end{gather*}
In this section we denote the spectral parameters by $z_j =x^{2u_j}$,
and denote integral variables by $w_a =x^{2v_a}$.

By using the vertex-face transformation,
we can rewrite (\ref{eq:d-ff}) as follows:
\begin{gather*}
F^{(i)}_{m}(\sigma^{z}; u_1 , \dots , u_{2m})_{
\nu_{1} \cdots \nu_{2m}}
= \dfrac{1}{\chi^{(i)}} \sum_{l_1,\dots ,
l_{2m}}\!\! t'{}^*_{\nu_1}
\left(u_{1}-u_0+\tfrac{1}{2}\right){}_{l}^{l_1}
\cdots t'{}^*_{\nu_{2m}}
\left(u_{2m}-u_0+\tfrac{1}{2}\right){}_{
l_{2m-1}}^{l_{2m}} \\
\qquad{}\times  \sum_{
k\equiv l+i\mbox{\scriptsize (mod $2$)}}\sum_{
\varepsilon =\pm}\varepsilon
\sum_{k_1=k\pm1}\sum_{k_2 =k_1\pm 1} t^*_{\varepsilon}(u-u_0)^{k}_{k_1}
t^{\varepsilon}(u-u_0)^{k_1}_{k_2} \\
\qquad{} \times  \mbox{Tr}_{{\cal H}^{(i)}_{l,k}}
\left( \Psi^* (u_1)^{l}_{l_{1}}\cdots \Psi^* (u_{2m})^{l_{2m-1}}_{l_{2m}}
\Phi^* (u)^{k}_{k_1} \Phi (u)^{k_{1}}_{k_2}
\Lambda (u_0 )_{l\,k}^{l_{2m} k_2} \dfrac{[k]x^{4H_F}}{[l]'} \right),
\end{gather*}
where $H_F$ is the CTM Hamiltonian def\/ined by (\ref{eq:CTM-Fock}).

Let
\begin{gather}
F^{(i)}_m(\sigma^{z}; u_1 , \dots , u_{2m})_{
ll_{1} \cdots l_{2m}}
= \dfrac{1}{\chi^{(i)}}  \sum_{
k\equiv l+i\, \mbox{\scriptsize (mod $2$)}}\sum_{
\varepsilon =\pm}\varepsilon
\sum_{k_1=k\pm1}\sum_{k_2 =k_1\pm 1}\! t^*_{\varepsilon}(u-u_0)^{k}_{k_1}
t^{\varepsilon}(u-u_0)^{k_1}_{k_2} \nonumber\\
\qquad{} \times  \mbox{Tr}_{{\cal H}^{(i)}_{l,k}}
\left( \Psi^* (u_1)^{l}_{l_{1}}\cdots \Psi^* (u_{2m})^{l_{2m-1}}_{l_{2m}}
\Phi^* (u)^{k}_{k_1} \Phi (u)^{k_{1}}_{k_2}
\Lambda (u_0 )_{l\,k}^{l_{2m} k_2} \dfrac{[k]x^{4H_F}}{[l]'} \right).
\label{eq:ff-tilde}
\end{gather}
Then we have
\begin{gather*}
F^{(i)}_m(\sigma^{z}; u_1 , \dots , u_{2m})_{
ll_{1} \cdots l_{2m}}  = \sum_{\nu_1, \dots , \nu_{2m}}
F^{(i)}_{m}(\sigma^{z}; u_1 , \dots , u_{2m})_{
\nu_{1} \cdots \nu_{2m}} \\
\qquad{} \times
t'{}^{\nu_{1}}
\left(u_{1}-u_0+\tfrac{1}{2}\right){}^{
l}_{l_{1}}\cdots t'{}^{\nu_{2m}}
\left(u_{2m}-u_0+\tfrac{1}{2}\right){}^{l_{2m-1}}_{l_{2m}}.
\end{gather*}

For simplicity, let $l_j=l-j$ for $1\leqslant j\leqslant 2m$. Then from
the relation (\ref{eq:Lamda-delta}),
$\Lambda (u_0 )_{l\,k}^{l_{2m} k_2}$ vanishes unless $k_2 =k-2$.
Thus, the sum over $k_1$ and $k_2$ on (\ref{eq:ff-tilde})
reduces to only one non-vanishing term. Furthermore, we note
the formula
\begin{gather*}
\sum_{
\varepsilon =\pm} \varepsilon
 t^*_{\varepsilon}(u-u_0)^{k}_{k-1}
t^{\varepsilon}(u-u_0)^{k-1}_{k-2}=(-1)^{1-i}\dfrac{
\{0\}\{u-u_0-1+k\}}{[u-u_0][k-1]}.
\end{gather*}
Here, we use $k-l\equiv i$ (mod $2$). The sum with respect to $k$
for the trace over the zero-modes parts
can be calculated as follows:
\begin{gather*}
\sum_{k\equiv l+i\, \mbox{\scriptsize (mod $2$)}} \{ u-u_0-1+k\}
\prod_{j=1}^{2m} (-z_j)^{\frac{rl}{2(r-1)}-\frac{k}{2}}
(x^{-1}z)^{-l+\frac{(r-1)k}{r}}   \prod_{a=1}^{m-1}
(-w_a)^{-\frac{rl}{r-1}+k} \\
\qquad\quad{} \times
\left( (-1)^r x^{-r+1}z_0 \right)^{-\frac{l}{r-1}+\frac{k}{r}}
x^{\frac{rl^2}{r-1}-2kl+\frac{(r-1)k^2}{r}} \\
\qquad {} =x^{\frac{l^2}{r-1}+l\Big(2+\frac{r}{r-1}\sum\limits_{j=1}^{2n} u_j -2u-
\frac{2r}{r-1}\sum\limits_{a=1}^{m-1} v_a -\frac{2u_0}{r-1}\Big)}
x^{\frac{1}{r}(u-u_0 -1)^2-(u-u_0-1)}
\sum_{k\equiv l+i\, \mbox{\scriptsize (mod $2$)}}
x^{(k-l)^2} \\
\qquad\quad{} \times \sum_{n\in\mathbb{Z}} x^{rn(n-1)}
x^{2(u-u_0-1+k)n} x^{k\Big( 2\sum\limits_{a=1}^{m-1}v_a +2u -
\sum\limits_{j=1}^{2m} u_j -3\Big)} \\
\qquad{} = \frac{(-1)^{1-i}}{2} x^{\frac{1}{r}(u-u_0 -1)^2-\frac{1}{r-1}
\Big(u_0 +\sum\limits_{a=1}^{m-1} v_a -\frac{1}{2}\sum\limits_{j=1}^{2m} u_j\Big)^2
-\Big(\sum\limits_{a=1}^{m-1} v_a +u-\frac{1}{2}\sum\limits_{j=1}^{2m} u_j -1\Big)^2}
\\ \qquad\quad{}
\times Z^{(i)}_m (l, u, u_0, u_j , v_a),
\end{gather*}
where
\begin{gather*}
Z^{(i)}_m (l, u, u_0, u_j , v_a)  =
\left[ l-u_0 - \sum_{a=1}^{m-1} v_a
+\tfrac{1}{2}\sum_{j=1}^{2m} u_j \right]'
\left[  \sum_{a=1}^{m-1} v_a +u
-\tfrac{1}{2}\sum_{j=1}^{2m} u_j\right]_1 \\
\qquad{} +(-1)^{1-i} \left\{ l-u_0 - \sum_{a=1}^{m-1} v_a
+\tfrac{1}{2}\sum_{j=1}^{2m} u_j \right\}'
\left\{ \sum_{a=1}^{m-1} v_a +u
-\tfrac{1}{2}\sum_{j=1}^{2m} u_j\right\}_1 .
\end{gather*}
Thus,
$F^{(i)}_m(\sigma^{z}; u_1 , \dots , u_{2m})_{
ll-1\cdots l-2m}$ can be obtained as follows:
\begin{gather}
(-1)^{m-1}\beta_m^{-1}F^{(i)}_m(\sigma^{z}; u_1 , \dots , u_{2m})_{
ll-1\cdots l-2m} \nonumber\\
= \prod_{j<j'} (-z_{j})^{\frac{r}{2(r-1)}}
F_{\psi^*\psi^*}(z_{j'}/z_j)
\prod_{j=1}^{2m} \dfrac{(-z_j)^{-\frac{1}{r-1}}
x^{\frac{(u_j-u_0+1/2)^2}{4(r-1)}+\frac{r(u_j-u_0+1/2)}{2(r-1)}
+\frac{1}{4}}}{x^{-1}z(xz_j/z; x^4)_\infty
(x^{3}z/z_j; x^4)_\infty}
f'(u_j-u_0+\tfrac{1}{2}) \nonumber\\
\times \oint_{C}
\prod_{a=1}^{m-1} \dfrac{dw_a}{2\pi\sqrt{-1}w_a}
Z^{(i)}_m (l, u, u_0, u_j , v_a)
\prod_{a<b} (-w_b)^{\frac{2r}{r-1}} [v_a -v_b]'
[v_a -v_b]_1 x^{-\frac{r}{r-1}(v_a-v_b-1)^2} \nonumber\\
\times \prod_{a=1}^{m-1} x^{-2}z^2
x^{-(v_a -u)^2+v_a-u}[v_a -u]_1 (-w_a)^{\frac{2}{r-1}}
x^{-\frac{1}{r-1}(u_0-v_a-1)^2+u_0-v_a-1}
[v_a-u_0+l-m]' \nonumber\\
\times \prod_{a=1}^{m-1}\prod_{j=1}^{2m}
(-z_j)^{-\frac{r}{r-1}} \dfrac{
(x^{2r-1}w_a/z_j; x^4, x^{2r-2})_\infty
(x^{2r+3}z_j/w_a; x^4, x^{2r-2})_\infty}{
(x^{-1}w_a/z_j; x^4, x^{2r-2})_\infty
(x^{3}z_j/w_a; x^4, x^{2r-2})_\infty} \nonumber\\
\times \dfrac{(x^{-1}z)^{\frac{2}{r}}}{2}
x^{-\frac{r+2}{r}(u_0-u)-\frac{1}{r}
-\frac{1}{r-1}
\Big(u_0 +\sum\limits_{a=1}^{m-1} v_a -\frac{1}{2}\sum\limits_{j=1}^{2m} u_j\Big)^2
-\Big(\sum\limits_{a=1}^{m-1} v_a +u-\frac{1}{2}\sum\limits_{j=1}^{2m} u_j -1\Big)^2},
\label{eq:F_l..l-2m}
\end{gather}
where $f'(v)$ is def\/ined by
(\ref{eq:f'-def}) for $n=2$, a scalar function $F_{\psi^*\psi^*}(z)$
and a scalar $\beta_m$ are
\begin{gather*}
F_{\psi^*\psi^*}(z)
=\dfrac{(z; x^4, x^4, x^{2r-2})_\infty
(x^4z^{-1}; x^4, x^4, x^{2r-2})_\infty}
{(x^2z; x^4, x^4, x^{2r-2})_\infty
(x^6z^{-1}; x^4, x^4, x^{2r-2})_\infty}
\\
\phantom{F_{\psi^*\psi^*}(z)=}{} \times\dfrac{(x^{2r+2}z; x^4, x^4, x^{2r-2})_\infty
(x^{2r+6}z^{-1}; x^4, x^4, x^{2r-2})_\infty}
{(x^{2r}z; x^4, x^4, x^{2r-2})_\infty
(x^{2r+4}z^{-1}; x^4, x^4, x^{2r-2})_\infty},
\end{gather*}
and
\begin{gather*}
\beta_m  = \dfrac{x^{-\frac{r-1}{4r}} \{ 0\} [m-1]'!
(x^{-2}z)^{\frac{r-1}{2r}}(x^2, x^4)_\infty^2
(x^2; x^{2r})_\infty (x^{2r+1}; x^{2r-2})_\infty}
{(m-1)![1]'{}^m
(x^{-1}-x) g_{1}(x^2) (x^{2r}; x^{2r})_\infty^2
(x^{2r+1}; x^{2r})_\infty} \\
\phantom{\beta_m  =}{}  \times  (x^2; x^2)_\infty^{m-1}(x^{2r}; x^{2r-2})_\infty^{m-1}
\dfrac{
(x^4; x^4, x^4, x^{2r-2})_\infty^m
(x^{2r+6}; x^4, x^4, x^{2r-2})_\infty^m}
{(x^6; x^4, x^4, x^{2r-2})_\infty^m
(x^{2r+4}; x^4, x^4, x^{2r-2})_\infty^m},
\end{gather*}
with
\begin{gather*}
[m]'!=\prod_{p=1}^m [p]'.
\end{gather*}
On (\ref{eq:F_l..l-2m}), the integral contour $C$ should be
chosen such that all integral variables $w_a$ lie in the
convergence domain $x^3|z_j|<|w_a|<x|z_j|$.

Gathering phase factors on (\ref{eq:F_l..l-2m}),
we have $e^{-\pi\sqrt{-1}\frac{3mr}{2(r-1)}}$.
Redef\/ining $f'(v)$ by a scalar factor, we thus obtain
the equality:
\begin{gather}
\sum_{\nu_1, \dots , \nu_{2m}}
F^{(i)}_{m}(\sigma^{z}; u_1 , \dots , u_{2m})_{
\nu_{1} \cdots \nu_{2m}} \prod_{j=1}^{2m}
\vartheta\left[\begin{array}{c} 0 \\ b_{\nu_j} \end{array} \right]
\left(\tfrac{u_j-u_0+\frac{1}{2}+l-j+1}{2(r-1)};
\tfrac{\pi\sqrt{-1}}{2\epsilon (r-1)} \right) \nonumber\\
= \beta_m  \prod_{j<j'} z_{j}^{\frac{r}{2(r-1)}}
F_{\psi^*\psi^*}(z_{j'}/z_j)
\prod_{j=1}^{2m} \dfrac{z_j^{-\frac{1}{r-1}}
}{x^{-1}z(xz_j/z; x^4)_\infty
(x^{3}z/z_j; x^4)_\infty} \nonumber \\
\times \oint_{C}
\prod_{a=1}^{m-1} \dfrac{dw_a}{2\pi\sqrt{-1}w_a}
Z^{(i)}_m (l, u, u_0, u_j , v_a)
\prod_{a<b} w_b^{\frac{2r}{r-1}} [v_a -v_b]'
[v_a -v_b]_1 x^{-\frac{r}{r-1}(v_a-v_b-1)^2} \nonumber\\
\times \prod_{a=1}^{m-1} x^{-2}z^2
x^{-(v_a -u)^2+v_a-u}[v_a -u]_1 w_a^{\frac{2}{r-1}}
x^{-\frac{1}{r-1}(u_0-v_a-1)^2+u_0-v_a-1}
[v_a-u_0+l-m]' \nonumber\\
\times \prod_{a=1}^{m-1}\prod_{j=1}^{2m}
z_j^{-\frac{r}{r-1}} \dfrac{
(x^{2r-1}w_a/z_j; x^4, x^{2r-2})_\infty
(x^{2r+3}z_j/w_a; x^4, x^{2r-2})_\infty}{
(x^{-1}w_a/z_j; x^4, x^{2r-2})_\infty
(x^{3}z_j/w_a; x^4, x^{2r-2})_\infty} \nonumber\\
\times \dfrac{(x^{-1}z)^{\frac{2}{r}}}{2}
x^{-\frac{r+2}{r}(u_0-u)-\frac{1}{r}
-\frac{1}{r-1}
\Big(u_0 +\sum\limits_{a=1}^{m-1} v_a -\frac{1}{2}\sum\limits_{j=1}^{2m} u_j\Big)^2
-\Big(\sum\limits_{a=1}^{m-1} v_a +u-\frac{1}{2}\sum\limits_{j=1}^{2m} u_j -1\Big)^2},
\label{eq:F_nu=F_l}
\end{gather}
where
\begin{gather*}
b_\nu =\left\{ \begin{array}{ll} 0 & (\nu =+), \\
\frac{1}{2} & (\nu =-).
\end{array} \right.
\end{gather*}
By comparing the transformation properties with respect to $l$
for both sides on (\ref{eq:F_nu=F_l}), we conclude that
$F^{(i)}_{m}(\sigma^{z}; u_1 , \dots , u_{2m})_{
\nu_{1} \cdots \nu_{2m}}$ are independent of $l$, and also that{\samepage
\begin{gather*}
\mbox{$F^{(i)}_{m}(\sigma^{z}; u_1 , \dots , u_{2m})_{
\nu_{1} \cdots \nu_{2m}}=0$ ~~~~ unless ~~
$\dfrac{1}{2}\displaystyle\sum_{j=1}^{2m} \nu_j \equiv 0
~~ \mbox{(mod $2$)}$},
\end{gather*}
as expected.}

When $m=1$, we have
\begin{gather}
F^{(i)}_{}(\sigma^{z}; u_1 , u_{2})_{\nu_{1} \nu_{2}} =
\delta_{\nu_{1}+\nu_{2},0}C^{(z)} z_{1}^{\frac{r}{2(r-1)}}
\displaystyle\prod_{j=1}^{2} \dfrac{z_j^{-\frac{1}{r-1}}
}{x^{-1}z(xz_j/z; x^4)_\infty
(x^{3}z/z_j; x^4)_\infty} \nonumber\\
\phantom{F^{(i)}_{}(\sigma^{z}; u_1 , u_{2})_{\nu_{1} \nu_{2}} = }{}
 \times \dfrac{(x^{-1}z)^{\frac{2}{r}}}{4}
x^{-\frac{r+2}{r}(u_0-u)-\frac{1}{r}
-\frac{1}{r-1}
(u_0 -(u_1+u_2)/2)^2
-(u-(u_1+u_2)/2 -1)^2}
\nonumber\\
\phantom{F^{(i)}_{}(\sigma^{z}; u_1 , u_{2})_{\nu_{1} \nu_{2}} = }{}
\times
F_{\psi^*\psi^*}(z_{2}/z_1) \left( \nu_1
\frac{[u-\frac{u_1+u_2}{2}]_1}{[\frac{u_2-u_1-1}{2}]'}
+(-1)^{1-i} \frac{\{ u-\frac{u_1+u_2}{2}\}_1}{\{\frac{u_2-u_1-1}{2}\}'}
\right),
\label{eq:sigma-z;m=1}
\end{gather}
where $C^{(z)}$ is a constant.
This is a same result obtained
by Lashkevich in \cite{La}, up to a scalar factor\footnote{This scalar factor results from the dif\/ference between
the present normalization of the type II vertex operators and
that used in~\cite{La}.}.

Next, let us choose $\sigma^x$ as a local operator:
\begin{gather*}
\sigma^{z}=E^{(1)}_{+-}+E^{(1)}_{-+}.
\end{gather*}
Then the relation
for $F^{(i)}_{m}(\sigma^{x}; u_1 , \dots , u_{2m})_{
\nu_{1} \cdots \nu_{2m}}$ reduces to (\ref{eq:F_nu=F_l})
with $Z^{(i)}_m (l, u, u_0, u_j , v_a)$ replaced by
\begin{gather*}
X^{(i)}_m (l, u, u_0, u_j , v_a)   =
\left[\!\!\left[ l-u_0 -\displaystyle\sum_{a=1}^{m-1} v_a
+\tfrac{1}{2}\sum_{j=1}^{2m} u_j \right]\!\!\right]'
\left\{\!\!\!\!\left\{ \displaystyle\sum_{a=1}^{m-1} v_a +u
-\tfrac{1}{2}\sum_{j=1}^{2m} u_j\right\}\!\!\!\!\right\}_1 \\
\qquad{} + (-1)^{1-i}
 \left\{\!\!\!\!\left\{ l-u_0 -\displaystyle\sum_{a=1}^{m-1} v_a
+\tfrac{1}{2}\sum_{j=1}^{2m} u_j \right\}\!\!\!\!\right\}'
\left[\!\!\left[ \displaystyle\sum_{a=1}^{m-1} v_a +u
-\tfrac{1}{2}\sum_{j=1}^{2m} u_j\right]\!\!\right]_1 ,
\end{gather*}
and with $\{ 0\}$ in $\beta_m$ replaced by $[\![ 0]\!]$, respectively.
Here,
\begin{alignat*}{4}
& [\![ v]\!]:=x^{\frac{v^2}{r}}\Theta_{x^{2r}}(x^{2v+r}),
\qquad &&
[\![ v]\!]':=[\![ v]\!]|_{r\mapsto r-1},\qquad &&
[\![ v]\!]_1:=[\![ v]\!]|_{r\mapsto 1}, & \\
& \{\!\!\{ v\}\!\!\}:=x^{\frac{v^2}{r}}\Theta_{x^{2r}}(-x^{2v+r}), \qquad &&
\{\!\!\{ v\}\!\!\}':=\{\!\!\{ v \}\!\!\}|_{r\mapsto r-1}, \qquad &&
\{\!\!\{ v\}\!\!\}_1:=\{\!\!\{ v\}\!\!\}|_{r\mapsto 1}. &
\end{alignat*}
The transformation properties with respect to $l$ implies
that
$F^{(i)}_{m}(\sigma^{x}; u_1 , \dots , u_{2m})_{
\nu_{1} \cdots \nu_{2m}}$ are independent of $l$, and also that
\begin{gather*}
\mbox{$F^{(i)}_{m}(\sigma^{x}; u_1 , \dots , u_{2m})_{
\nu_{1} \cdots \nu_{2m}}=0$ ~~~~ unless ~~
$\dfrac{1}{2}\displaystyle\sum_{j=1}^{2m} \nu_j \equiv 1 ~~
\mbox{(mod $2$)}$},
\end{gather*}
as expected.
Furthermore, $2$-point form factors for $\sigma^x$-operator can be
obtained as follows:
\begin{gather}
F^{(i)}_{}(\sigma^{x}; u_1 , u_{2})_{\nu_{1} \nu_{2}} =
\delta_{\nu_{1}\,\nu_{2}}C^{(x)} z_{1}^{\frac{r}{2(r-1)}}
\displaystyle\prod_{j=1}^{2} \dfrac{z_j^{-\frac{1}{r-1}}
}{x^{-1}z(xz_j/z; x^4)_\infty
(x^{3}z/z_j; x^4)_\infty} \nonumber\\
\phantom{F^{(i)}_{}(\sigma^{x}; u_1 , u_{2})_{\nu_{1} \nu_{2}} = }{}
\times \dfrac{(x^{-1}z)^{\frac{2}{r}}}{4}
x^{-\frac{r+2}{r}(u_0-u)-\frac{1}{r}
-\frac{1}{r-1}
(u_0 -(u_1+u_2)/2)^2
-(u-(u_1+u_2)/2 -1)^2} \nonumber\\
\phantom{F^{(i)}_{}(\sigma^{x}; u_1 , u_{2})_{\nu_{1} \nu_{2}} = }{}
\times
F_{\psi^*\psi^*}(z_{2}/z_1)
\left( \nu_1
\frac{\{\!\!\{u-\frac{u_1+u_2}{2}\}\!\!\}_1}{[\![\frac{u_2-u_1-1}{2}\!]]'}
+(-1)^{1-i}
\frac{[\![ u-\frac{u_1+u_2}{2}\!]]_1}{\{\!\!\{\frac{u_2-u_1-1}{2}\}\!\!\}'}
\right),
\label{eq:sigma-x;m=1}
\end{gather}
where $C^{(x)}$ is a constant. The expressions (\ref{eq:sigma-z;m=1})
and (\ref{eq:sigma-x;m=1}) are essentially same as the results
obtained by Lukyanov and Terras \cite{LT}\footnote{Strictly speaking, we consider the parameterization of the coupling
constants $|J_z|>J_x>J_y$ while
Lukyanov and Terras \cite{LT} considered that of $J_x>J_y>|J_z|$.
Thus, the present results~(\ref{eq:sigma-z;m=1})
and~(\ref{eq:sigma-x;m=1}) correspond to their results of
the $2$-point form factors
for~$\sigma^x$-operator and $\sigma^y$-operator, respectively. Furthermore,
we note that their rapidity $\theta_j$ can be obtained from our
spectral parameter $u_j$ by a constant shift. After such substitution,
we claim that our results~(\ref{eq:sigma-z;m=1})
and~(\ref{eq:sigma-x;m=1}) agree with their corresponding results
in~\cite{LT}.}.

\section{Concluding remarks}\label{section6}

In this paper we present a vertex operator approach for form factors
of the $(\mathbb{Z}/n\mathbb{Z})$-symmetric model. For that purpose
we constructed the free f\/ield representations of the tail operators
$\Lambda_{\xi\,a}^{\xi' a'}$, the nonlocal operators which relate
the physical quantities of the $(\mathbb{Z}/n\mathbb{Z})$-symmetric model
and the $A^{(1)}_{n-1}$ model. As a result, we can obtain
the integral formulae for form factors
of the $(\mathbb{Z}/n\mathbb{Z})$-symmetric model, in principle.

Our approach is based on some assumptions. We assumed that the vertex
operator al\-gebra def\/ined by (\ref{eq:rho-rel}) and
(\ref{eq:Lambda-phi}), (\ref{eq:Lambda-psi}) correctly describes
the intertwining relation between the $(\mathbb{Z}/n\mathbb{Z})$-symmetric
model and the $A^{(1)}_{n-1}$ model. We also assumed that
the free f\/ield representations~(\ref{eq:Bose-Lambda}),~(\ref{eq:Lambda-repII}) provide
relevant representations of the vertex operator algebra.

As a consistency check of our bosonization scheme,
we presented the integral formulae for form factors which are related to
the $\sigma^z$-operator and $\sigma^x$-operator
in the eight-vertex model, i.e., the
$(\mathbb{Z}/2\mathbb{Z})$-symmetric model.
The expressions (\ref{eq:F_l..l-2m}) and (\ref{eq:F_nu=F_l}) for
$\sigma^z$ form factors and $\sigma^x$ analogues remind us
of the determinant structure of sine-Gordon form factors found by
Smirnov~\cite{Smbk}. In Smirnov's approach form factors in
integrable models can be obtained by solving matrix Riemann-Hilbert
problems. We wish to f\/ind form factors formulae in
the eight-vertex model on the basis of Smirnov's approach
in a separate paper.

\appendix

\section{OPE formulae and commutation relations}\label{AppendixA}

In this paper we use some dif\/ferent def\/initions of the
basic bosons from the one used in \cite{Bel-form}.
Accordingly, some formulae listed in Appendix B of \cite{Bel-form}
should be changed. Here we list such formulae. Concerning unchanged
formulae see \cite{Bel-form}.
In what follows we denote
$z=x^{2v}$, $z'=x^{2v'}$.

First, useful OPE formulae are:
\begin{gather}
V_{\omega_1}(v)V_{\omega_j}(v') =  (-z)^{
\frac{r}{r-1}\frac{n-j}{n}} g^*_j(z'/z)
:V_{\omega_1}(v)V_{\omega_j}(v'):, \nonumber\\
V_{\omega_j}(v)V_{\omega_1}(v') =  (-z)^{
\frac{r}{r-1}\frac{n-j}{n}} g^*_j(z'/z)
:V_{\omega_j}(v)V_{\omega_1}(v'):, \nonumber\\\
V_{\omega_j}(v)V_{-\alpha_j}(v') = (-z)^{-\frac{r}{r-1}}
\dfrac{(x^{2r-1}z'/z; x^{2r-2})_\infty}{(x^{-1}z'/z; x^{2r-2})_\infty}
:V_{\omega_j}(v)V_{-\alpha_j}(v'):,\nonumber \\ 
V_{-\alpha_j}(v)V_{\omega_j}(v') = (-z)^{-\frac{r}{r-1}}
\dfrac{(x^{2r-1}z'/z; x^{2r-2})_\infty}{(x^{-1}z'/z; x^{2r-2})_\infty}
:V_{-\alpha_j}(v)V_{\omega_j}(v'):, \nonumber\\
V_{-\alpha_{j}}(v)V_{-\alpha_{j\pm 1}}(v') = (-z)^{-\frac{r}{r-1}}
\dfrac{(x^{2r-1}z'/z; x^{2r-2})_\infty}{(x^{-1}z'/z; x^{2r-2})_\infty}
:V_{-\alpha_{j}}(v)V_{-\alpha_{j\pm 1}}(v'):, \nonumber\\ 
V_{-\alpha_{j}}(v)V_{-\alpha_{j}}(v') = (-z)^{\frac{2r}{r-1}} \left(
1-\dfrac{z'}{z}\right)
\dfrac{(x^{-2}z'/z; x^{2r-2})_\infty}{(x^{2r}z'/z; x^{2r-2})_\infty}
:V_{-\alpha_{j}}(v)V_{-\alpha_{j}}(v'):, \nonumber\\
V_{\omega_j}(v)U_{\omega_j}(v') =  (-z)^{-\frac{j(n-j)}{n}} \rho_j (z'/z)
:V_{\omega_1}(v)U_{\omega_j}(v'):, \nonumber\\
U_{\omega_j}(v)V_{\omega_j}(v') =  z^{-\frac{j(n-j)}{n}} \rho_j (z'/z)
:U_{\omega_j}(v)V_{\omega_j}(v'):, \nonumber\\
V_{\omega_{j}}(v)U_{-\alpha_{j}}(v') = -z\left( 1-\dfrac{z'}{z} \right)
:V_{\omega_{j}}(v)U_{-\alpha_{j}}(v'):  =
U_{-\alpha_{j}}(v')V_{\omega_{j}}(v), \nonumber\\ 
U_{\omega_{j}}(v)V_{-\alpha_{j}}(v') = z\left( 1-\dfrac{z'}{z} \right)
:U_{\omega_{j}}(v)V_{-\alpha_{j}}(v'):  =
V_{-\alpha_{j}}(v')U_{\omega_{j}}(v), \nonumber\\
V_{-\alpha_{j}}(v)U_{-\alpha_{j\pm 1}}(v') = -z\left( 1-\dfrac{z'}{z} \right)
:V_{-\alpha_{j}}(v)U_{-\alpha_{j\pm 1}}(v'):  =
U_{-\alpha_{j\pm 1}}(v')V_{-\alpha_{j}}(v), \nonumber\\ 
V_{-\alpha_{j}}(v)U_{-\alpha_{j}}(v') =
\dfrac{:V_{-\alpha_{j}}(v)U_{-\alpha_{j}}(v'):}{
z^2(1-\frac{xz'}{z})(1-\frac{x^{-1}z'}{z})}, \label{eq:VU-normal} \\
U_{-\alpha_{j}}(v)V_{-\alpha_{j}}(v') =
\dfrac{:U_{-\alpha_{j}}(v)V_{-\alpha_{j}}(v'):}{
z{}^2(1-\frac{xz'}{z})(1-\frac{x^{-1}z'}{z})}, \label{eq:UV-normal}
\end{gather}
where $g^*_j (z)$ and $\rho_j (z)$
are def\/ined by (\ref{eq:g*-def}) and
(\ref{eq:chi-def}), respectively. From (\ref{eq:VU-normal}) and
(\ref{eq:UV-normal}), we obtain the following
commutation relations:
\begin{gather*}
 [ V_{-\alpha_{j}}(v), U_{-\alpha_{j}}(v') ] =
\dfrac{\delta (\frac{z}{xz'})-\delta (\frac{z'}{xz})}{
(x-x^{-1})zz'}
:V_{-\alpha_{j}}(v)U_{-\alpha_{j}}(v'):,
\end{gather*}
where
the $\delta$-function is def\/ined by the following formal
power series
\begin{gather*}
\delta (z)=\sum_{n\in \mathbb{Z}} z^n.
\end{gather*}

Finally, we list the OPE formulae for $W_{-\alpha_j}(v)$ and
other basic operators:
\begin{gather*}
W_{-\alpha_{j}}(v)V_{-\alpha_{j\pm 1}}(v') =
-(-z)^{-\frac{1}{r-1}} \dfrac{(x^{r}z'/z; x^{2r-2})_\infty}
{(x^{r-2}z'/z; x^{2r-2})_\infty}
:W_{-\alpha_{j}}(v)V_{-\alpha_{j\pm 1}}(v'):, \\ 
V_{-\alpha_{j\pm 1}}(v)W_{-\alpha_{j}}(v') =
(-z)^{-\frac{1}{r-1}} \dfrac{(x^{r}z'/z; x^{2r-2})_\infty}
{(x^{r-2}z'/z; x^{2r-2})_\infty}
:V_{-\alpha_{j\pm 1}}(v)W_{-\alpha_{j}}(v'):, \\ 
V_{\omega_j}(v)W_{-\alpha_{j}}(v') =
(-z)^{-\frac{1}{r-1}} \dfrac{(x^{r}z'/z; x^{2r-2})_\infty}
{(x^{r-2}z'/z; x^{2r-2})_\infty}
:V_{\omega_j}(v)W_{-\alpha_{j}}(v'):, \\ 
W_{-\alpha_{j}}(v)V_{\omega_j}(v') =
-(-z)^{-\frac{1}{r-1}} \dfrac{(x^{r}z'/z; x^{2r-2})_\infty}
{(x^{r-2}z'/z; x^{2r-2})_\infty}
:W_{-\alpha_{j}}(v)V_{\omega_j}(v'):, 
\end{gather*}
From these, we obtain
\begin{gather*}
W_{-\alpha_j}\left( v+\tfrac{r}{2}
\right) V_{-\alpha_{j\pm 1}}(v) = 0 =
V_{-\alpha_{j\pm 1}}(v)
W_{-\alpha_j}\left( v-\tfrac{r}{2}
\right) , \\ 
W_{-\alpha_j}\left( v+\tfrac{r}{2}
\right) V_{\omega_j}(v)
 = 0 = V_{\omega_j}(v)
W_{-\alpha_j}\left( v-\tfrac{r}{2}
\right). 
\end{gather*}

\subsection*{Acknowledgements}

We would like to thank T.~Deguchi, R.~Inoue, H.~Konno, Y.~Takeyama
and R.~Weston for discussion and their interests in the present work.
We would also like to thank S.~Lukyanov for useful information.
This paper is partly based on a talk given
in International Workshop RAQIS'10,
Recent Advances in Quantum Integrable Systems,
held at LAPTH, Annecy-le-Vieux, France,  June~15--18, 2010.
We would like to thank L.~Frappat and \'{E}.~Ragoucy
for organizing the conference.

\pdfbookmark[1]{References}{ref}
\LastPageEnding

\end{document}